\documentclass[twocolumn,prl,showkeys,superscriptaddress,groupedaddress]{revtex4}
\usepackage{epsfig,color}
\usepackage{amssymb}
\usepackage{amsmath}
\usepackage{dcolumn}
\usepackage{graphicx}

\usepackage[pdfborder={0 0 0}, colorlinks=true, linkcolor=blue, citecolor=blue]{hyperref}

\def\beq{\begin{equation}}
\def\eeq{\end{equation}}
\def\bea{\begin{eqnarray}}
\def\eea{\end{eqnarray}}

\def\q{\mathbf{q}}
\def\k{\mathbf{k}}
\def\p{\mathbf{p}}

\def\Q{\mathbf{Q}}

\newcommand{\ket}[1]{\left|{#1}\right>}

\newcommand{\eps}{\varepsilon}

\newcommand{\su}{\uparrow}
\newcommand{\sd}{\downarrow}

\newcommand{\sign}{\mathrm{sgn}}

\newcommand{\ii}{{\mathrm{i}}}

\newcommand{\nn}{\nonumber}

\begin{document}

\title{Role of interactions in the energy of the spin resonance peak in Fe-based superconductors}

\author{Yu.N.~Togushova}
\affiliation{Siberian Federal University, Svobodny Prospect 79, 660041 Krasnoyarsk, Russia}

\author{M.M.~Korshunov}
\email{mkor@iph.krasn.ru}
\affiliation{Siberian Federal University, Svobodny Prospect 79, 660041 Krasnoyarsk, Russia}
\affiliation{Kirensky Institute of Physics, Federal Research Center KSC SB RAS, 660036 Krasnoyarsk, Russia}

\date{\today}

\begin{abstract}
We consider the spin response within the five-orbital model for iron-based superconductors and study two cases: equal and unequal gaps in different bands. In the first case, the spin resonance peak in the superconducting state appears below the characteristic energy scale determined by the gap magnitude, $2\Delta_L$. In the second case, the energy scale corresponds to the sum of smaller and larger gap magnitudes, $\Delta_L + \Delta_S$. Increasing the values of the Hubbard interaction and the Hund's exchange, we observe a shift of the spin resonance energy to lower frequencies.
\end{abstract}


\maketitle

\section{Introduction}
%\vspace{-6pt}

Origin of the unconventional superconducting state in iron pnictides and chalcogenides is still under debate~\cite{Reviews}. Fe-based superconductors (FeBS) have square lattice of iron as the basic element, though with orthorhombic distortions in lightly doped materials. Iron is surrounded by As or P in pnictides or Se, Te, or S in chalcogenides.
Pnictides are represented by 1111 systems (LaFeAsO, LaFePO, Sr$_2$VO$_3$FeAs, etc.), 111 systems (LiFeAs, LiFeP, and others), and 122 systems (BaFe$_2$As$_2$, KFe$_2$As$_2$, and so on). Chalcogenides can be of 11 type (Fe$_{1-\delta}$Se, Fe$_{1+y}$Te$_{1-x}$Se$_x$, monolayers of FeSe) and of 122 type (KFe$_2$Se$_2$). Fermi surface (FS) is formed by Fe $d$-orbitals. Conductivity is provided by the iron layer, thus, the discussion of physics in terms of quasi two-dimensional system in most cases gives reasonable results~\cite{ROPPreview2011}. Excluding the cases of extreme hole and electron dopings, FS consists of two hole sheets around the $\Gamma=(0,0)$ point and two electron sheets around the $(\pi,0)$ and $(0,\pi)$ points in the two-dimensional Brillouin zone (BZ) corresponding to one Fe per unit cell. Nesting between these two groups of pockets leads to the enhanced antiferromagnetic fluctuations with the maximal scattering near the wave vector $\Q=(\pi,0)$ connecting hole and electron pockets.

Since different mechanisms of Cooper pairing result in different gap symmetries and structures~\cite{ROPPreview2011}, one can elucidate the superconducting mechanism by determining the gap structure. For example, the RPA-SF (random-phase approximation spin fluctuation) approach gives the extended $s$-wave gap that changes sign between hole and electron FS sheets ($s_{\pm}$ state) as the main instability for the wide range of dopings~\cite{Mazin_etal_splusminus,Graser2009,Kuroki2008,MaitiPRB,KorshunovUFN}. On the other hand, orbital fluctuations results in the order parameter with the sign-preserving $s_{++}$ symmetry~\cite{Kontani}.

One of the specific features of the $s_\pm$ state is the spin resonance peak in the dynamical spin susceptibility $\chi(\Q,\omega)$. Since $\Q$ connects Fermi sheets with different signs of $s_\pm$ gaps, the resonance condition for the interband susceptibility is fulfilled and the spin resonance peak is formed at a frequency $\omega_R$ below $2\Delta$ with $\Delta$ being the gap size~\cite{Korshunov2008,Maier,Maier2}. It was observed below $T_c$ at or around $\q = \Q$ in inelastic neutron scattering experiments on 1111, 122, and 11 systems~\cite{ChristiansonBKFA,Inosov2010,Argyriou2010,Lumsden2011,Dai2015}.

As is known from angle-resolved photoemission spectroscopy (ARPES) and recent measurements of gaps via Andreev spectroscopy, there are at least two distinct gaps present in 11, 122, and 1111 systems~\cite{Daghero2009,Tortello2010,Ponomarev2013,Abdel-Hafiez2014,Kuzmichev2016} and even three gaps in LiFeAs~\cite{Kuzmichev2012,Kuzmichev2013}. Larger gap ($\Delta_L$) is located at electron FS sheets and at the inner hole sheet, and the smaller gap ($\Delta_S$) is located at the outer hole FS~\cite{Ding2008,Evtushinsky2009}. Previously, we have found that in the case of unequal gaps on hole and electron pockets, the spin resonance frequency should appear below the characteristic energy scale, $\omega_R \leq \Delta_L+\Delta_S$~\cite{KorshunovPRB2016}. Comparison of experimental data on the peak frequency and gaps magnitudes leads to conclusion that in most cases the observed peak fulfills the condition and, therefore, indicates the $s_\pm$ gap structure~\cite{KorshunovPRB2016,KorshunovJMMM2017}. Here we study the how the changes in model parameters affect the spin resonance frequency $\omega_R$. In particular, we show that the increase of local Coulomb interactions leads to the decrease of $\omega_R$.

\section{Model and approach}

\begin{figure}[t]
\begin{center}
\includegraphics[width=0.8\linewidth]{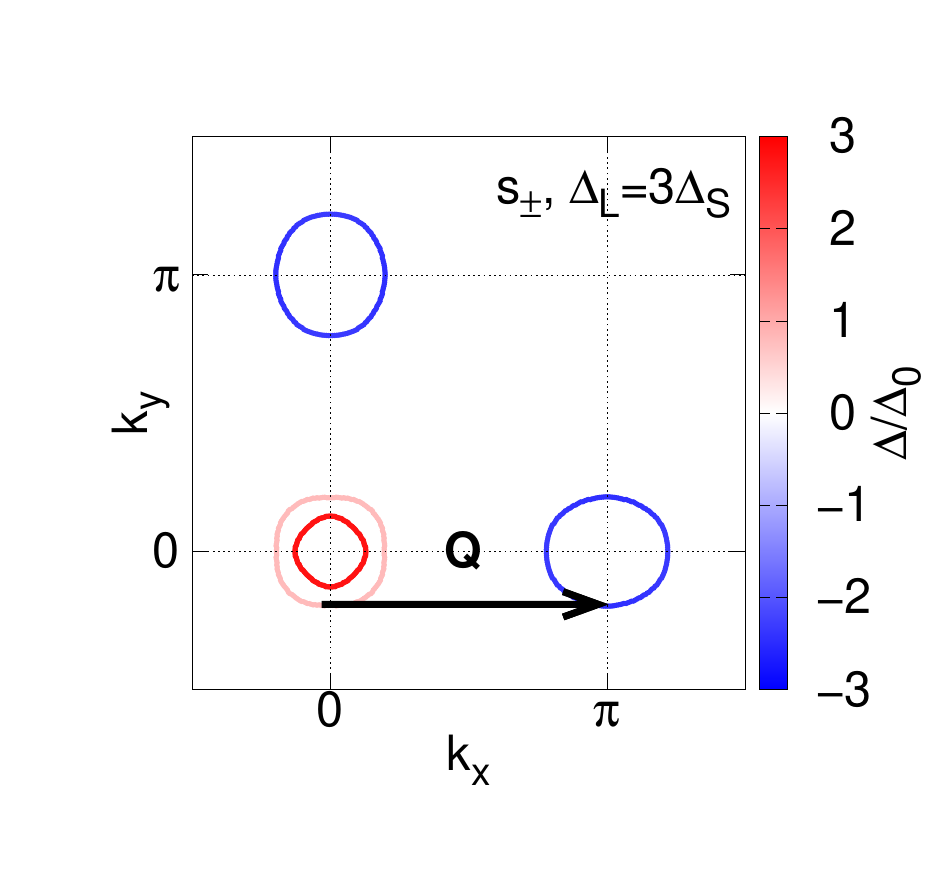}
\caption{Gaps at the Fermi surface for doping $x=0.05$ in the $s_\pm$ state with $\Delta_{\alpha_{1,2}} = \Delta_{\beta_{1}} = \Delta_L$ and $\Delta_{\beta_2} = \Delta_S$, where $\Delta_S = \Delta_L / 3$. Scattering wave vector $\Q$ entering the spin susceptibility is also shown. \label{fig:Delta}}
\end{center}
\end{figure}

To calculate spin susceptibility in normal and superconducting states, we use random phase approximation (RPA) with the local Coulomb interactions (Hubbard and Hund's exchange). In the multiorbital system, transverse dynamical spin susceptibility $\hat\chi_{+-}(\q,\omega)$ is the matrix in orbital indices. It can be obtained in the RPA from the bare electron-hole matrix bubble $\hat\chi_{(0)+-}(\q,\omega)$ by summing up a series of ladder diagrams,
\begin{equation}
 \hat\chi_{+-}(\q,\omega) = \left[\hat{I} - \hat{U}_s \hat\chi_{(0)+-}(\q,\omega)\right]^{-1} \hat\chi_{(0)+-}(\q,\omega),
 \label{eq:chi_s_sol}
\end{equation}
where $\q$ is the momentum, $\omega$ is the frequency, $\hat{U}_s$ and $\hat{I}$ are interaction and unit matrices in orbital space, respectively. Here we use the tight-binding model from Ref.~\cite{Graser2009} based on the fit to the DFT (density functional theory) band structure for prototypical pnictide LaFeAsO~\cite{Cao2008}. The model includes all five Fe $d$-orbitals and is given by
\begin{equation}
 H_0 = \sum_{\k \sigma} \sum_{l l'} \left[ t_{l l'}(\k) + \epsilon_{l} \delta_{l l'} \right] d_{l \k \sigma}^\dagger d_{l' \k \sigma},
 \label{eq:H0}
\end{equation}
where $d_{l \k \sigma}^\dagger$ is the annihilation operator of a particle with momentum $\k$, spin $\sigma$, and orbital index $l \in (1,2,\ldots,5)$ ($d_{xz}$, $d_{yz}$, $d_{xy}$, $d_{x^2-y^2}$, $d_{3z^2-r^2}$) . Later we use numerical values of hopping matrix elements $t_{l l'}(\k)$ and one-electron energies $\epsilon_{l}$ from Ref.~\cite{Graser2009}. This model for the undoped and moderately electron doped materials gives FS composed of two hole pockets, $\alpha_1$ and $\alpha_2$, around the $(0,0)$ point and two electron pockets, $\beta_1$ and $\beta_2$, centered around $(\pi,0)$ and $(0,\pi)$ points of the Brillouin zone. Total number of electrons is given by $n = n_0 \pm x$, where electron filling $n_0 = 6$ corresponds to the fully occupied $d^6$-orbital and $x$ is the doping concentration. Similar model for iron pnictides was proposed in Ref.~\cite{Kuroki2008}.

The general two-particle on-site Coulomb interaction is represented by the Hamiltonian~\cite{Graser2009,Kuroki2008,Castallani1978,Oles1983}:
\begin{eqnarray}
 H_{int} &=& U \sum_{f, m} n_{f m \su} n_{f m \sd} + U' \sum_{f, m < l} n_{f l} n_{f m} \nn\\
 &+& J \sum_{f, m < l} \sum_{\sigma,\sigma'} d_{f l \sigma}^\dag d_{f m \sigma'}^\dag d_{f l \sigma'} d_{f m \sigma}  \nn\\
 &+& J' \sum_{f, m \neq l} d_{f l \su}^\dag d_{f l \sd}^\dag d_{f m \sd} d_{f m \su}.
\label{eq:Hint}
\end{eqnarray}
where $n_{f m} = n_{f m \su} + n_{f m \sd}$, $n_{f m \sigma} = d_{f m \sigma}^\dag d_{f m \sigma}$ is the number of particles operator at the site $f$, $U$ and $U'$ are the intra- and interorbital Hubbard repulsion, $J$ is the Hund's exchange, and $J'$ is the so-called pair hopping.

Green's functions are diagonal in the band basis, $G_{\mu \sigma}(\k,\ii\omega_n) = 1 / \left( \ii\omega_n - \eps_{\k\mu\sigma} \right)$ with $\mu$ being the band index, but not in the orbital basis. Transformation from the orbital to the band basis is done via the matrix elements $\varphi^{\mu}_{\k m}$: $\ket{\sigma m \k} = \sum\limits_{\mu} \varphi^{\mu}_{\k m} \ket{\sigma \mu \k}$. Then $d_{\k m \sigma} = \sum\limits_{\mu} \varphi^{\mu}_{\k m} b_{\k \mu \sigma}$, where $b_{\k \mu \sigma}$ is the annihilation operator of electron. Transverse component of the bare spin susceptibility that is a tensor in orbital indices $l$, $l'$, $m$, and $m'$ takes the following form~\cite{KorshunovUFN},
\begin{widetext}
\begin{eqnarray}
 \label{eq:chipmmu}
 \chi^{ll',mm'}_{(0)+-}(\q,\ii\Omega) &=& -T \sum_{\p,\omega_n, \mu,\nu} \left[ \varphi^{\mu}_{\p m} {\varphi^*}^{\mu}_{\p l} G_{\mu \su}(p,\ii\omega_n) G_{\nu \sd}(\p+\q,\ii\Omega+\ii\omega_n) \varphi^{\nu}_{\p+\q l'} {\varphi^*}^{\nu}_{\p+\q m'} \right. \nn\\
 &-& \left.{\varphi^*}^{\mu}_{\p l} {\varphi^*}^{\mu}_{-\p m'} F^\dag_{\mu \su}(\p,-\ii\omega_n) F_{\nu \sd}(\p+\q,\ii\Omega+\ii\omega_n) \varphi^{\nu}_{\p+\q l'} \varphi^{\nu}_{-\p-\q m} \right].
\end{eqnarray}
\end{widetext}
Here $G_{\mu \su}(p,\ii\omega_n)$ and $F_{\mu \su}(\p,\ii\omega_n)$ are normal and anomalous (Gor'kov) Green's functions, $\ii\Omega$ is the Matsubara frequency.

Components of the physical spin susceptibility, $\chi_{+-}(\q,\ii\Omega) = \frac{1}{2} \sum_{l,m} \chi^{ll,mm}_{+-}(\q,\ii\Omega)$, are calculated using Eq.~(\ref{eq:chi_s_sol}) with the interaction matrix $U_s$ from Ref.~\cite{Graser2009}. To use matrix notations in Eq.~(\ref{eq:chi_s_sol}), we introduce the correspondence between matrix ($\imath$, $\jmath$) and orbital indices: $\imath = l + l' n_O$ and $\jmath = m + m' n_O$, where $n_O$ is the number of orbitals.

Since calculation of the Cooper pairing instability is not a topic of the present study, here we assume that the superconductivity is coming from some other theory and study the $s_\pm$ state with $\Delta_{\k\mu} = \Delta_{\mu} \cos k_x \cos k_y$, where $\mu$ is the band index. Two cases are considered below: equal gaps with $\Delta_{\mu'} = \Delta_{\mu}$ and unequal gaps with the smaller gap $\Delta_{\beta_2} = \Delta_S$ on the outer hole FS and larger gaps $\Delta_{\alpha_{1,2}} = \Delta_{\beta_{1}} = \Delta_L$ on inner hole and electron FSs. To be consistent with the experimental data, we choose $\Delta_S = \Delta_L / 3$, see Fig.~\ref{fig:Delta}.

\section{Results of calculations}

\begin{figure}[t]
\begin{center}
\includegraphics[width=1.\linewidth]{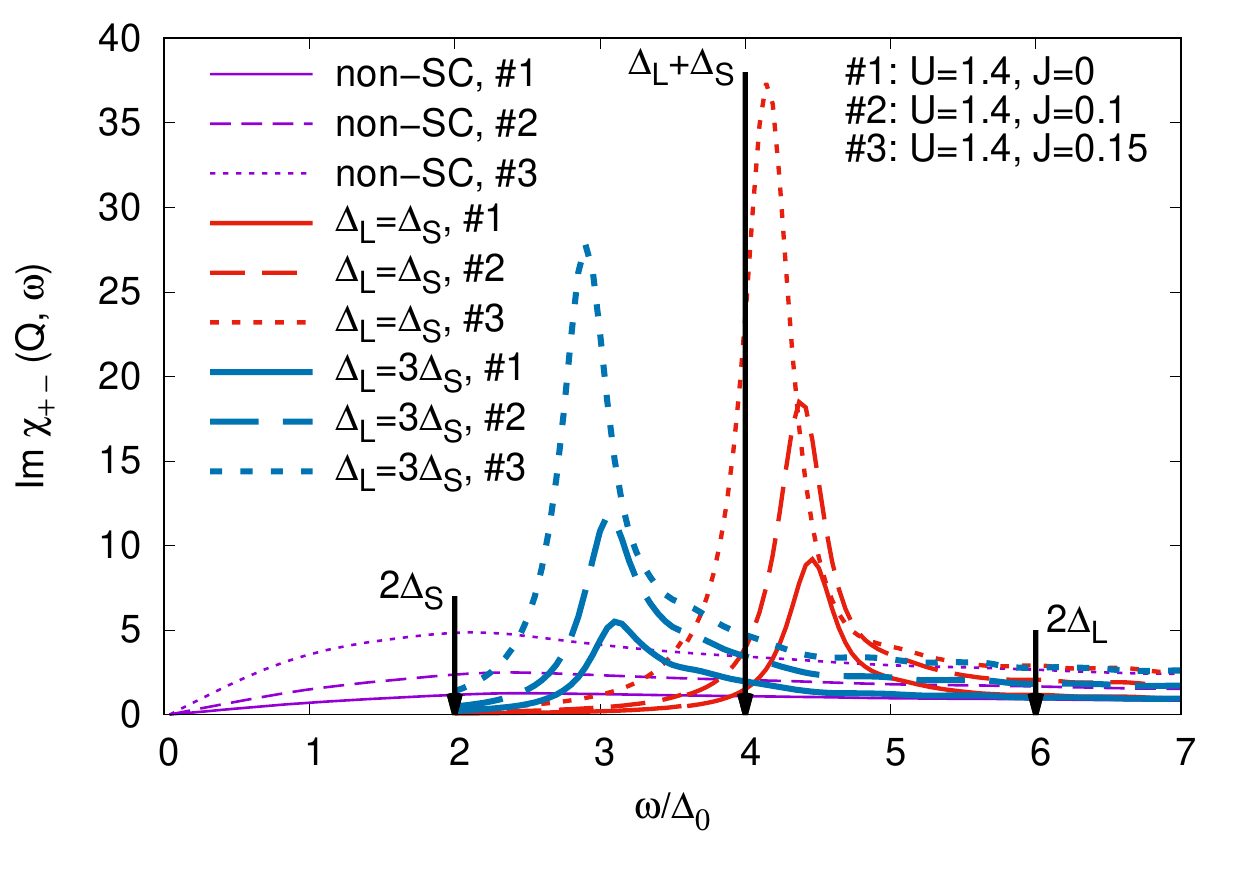}
\caption{Physical spin susceptibility $\mathrm{Im}\chi_{+-}(\Q,\omega)$ with $\Q=(\pi,0)$ for the five-orbital model in the normal (non-SC) and $s_\pm$ superconducting states. Two cases of superconducting states are shown: equal gaps with $\Delta_{\alpha_{1,2}} = \Delta_{\beta_{1,2}} = \Delta_L$, and unequal gaps with $\Delta_{\alpha_{1,2}} = \Delta_{\beta_{1}} = \Delta_L$ and $\Delta_{\beta_2} = \Delta_S$, where $\Delta_S = \Delta_L / 3$. All results are shown for the three sets of interaction parameters shown in figure (all values are in eV).
\label{fig:5orbImChi_set1}}
\end{center}
\end{figure}

\begin{figure}[t]
\begin{center}
\includegraphics[width=1.\linewidth]{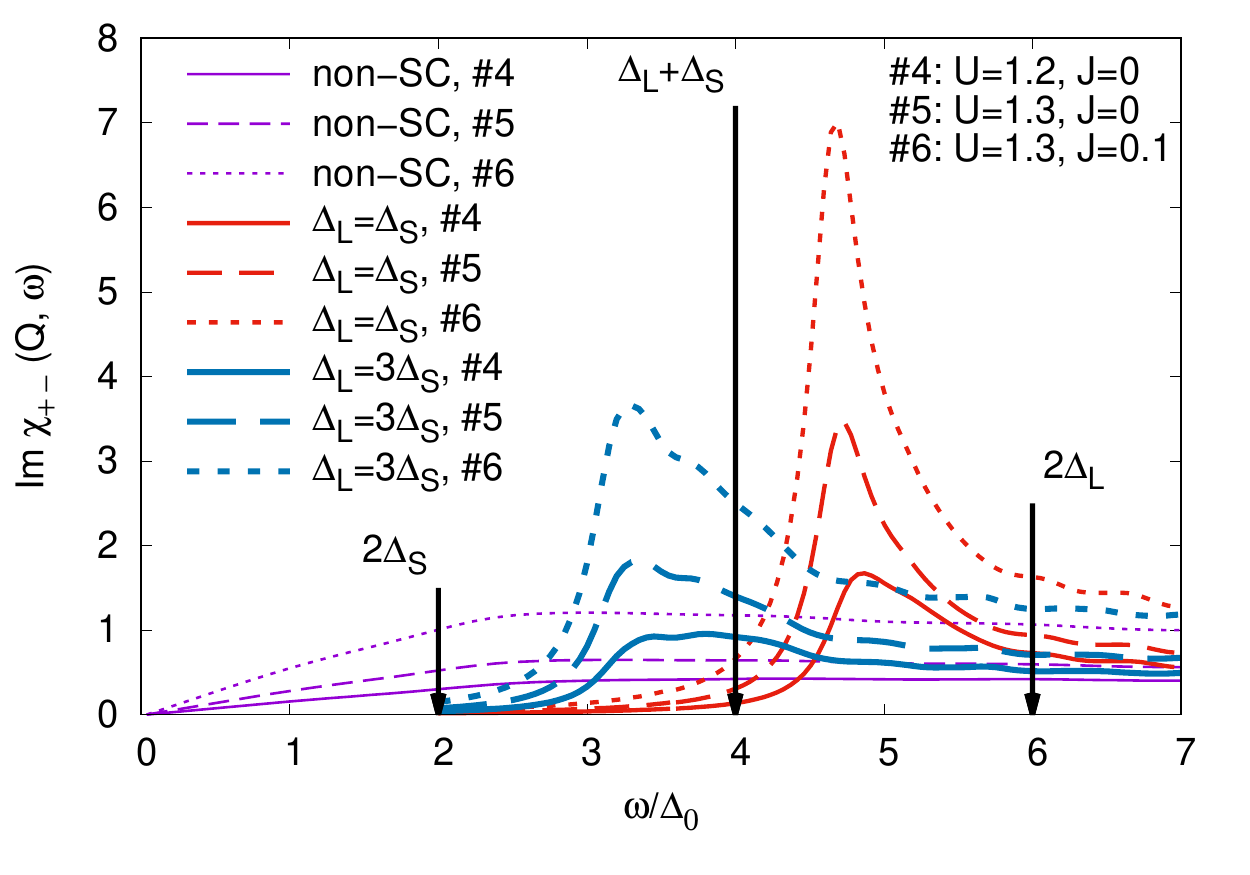}
\caption{The same, as in Fig.~\ref{fig:5orbImChi_set1}, but for different sets of interaction parameters. \label{fig:5orbImChi_set2}}
\end{center}
\end{figure}

In Figs.~\ref{fig:5orbImChi_set1} and~\ref{fig:5orbImChi_set2}, we present results for susceptibilities at the wave vector $\q=\Q$ with doping $x=0.05$ as functions of real frequency $\omega$ obtained via the analytical continuation from Matsubara frequencies ($\ii\Omega \to \omega + \ii\delta$ with $\delta \to 0+$). Since $\chi_{(0)+-}(\q,\omega)$ describes particle-hole excitations and in the superconducting state all excitations are gapped below approximately $2\Delta_0$ (at $T=0$), then $\mathrm{Im}\chi_{(0)+-}(\q,\omega)$ becomes finite only after that frequency. Due to the anomalous Green's functions, the anomalous coherence factors appear in~(\ref{eq:chipmmu}), which are proportional to $\left[1 - \frac{\Delta_{\k \mu} \Delta_{\k+\q \nu}}{E_{\k \mu} E_{\k+\q \nu}}\right]$ with $E_{\k \mu} \equiv \sqrt{\eps_{\k \mu}^2 + \Delta_{\k \mu}^2} = |\Delta_{\k \mu}|$ at the Fermi level.
%For the $s_{++}$ state, $\sign \Delta_{\k \mu} = \sign \Delta_{\k+\Q \nu}$, thus coherence factors vanish and there is a gradual increase of $\mathrm{Im}\chi_{(0)}$ for $\omega > 2\Delta_L$.
For the $s_\pm$ state, $\Q$ connects FSs with different signs of gaps, $\sign \Delta_{\k \mu} = - \sign \Delta_{\k+\Q \nu}$, thus coherence factors are finite and the imaginary part of $\chi_{(0)}$ possesses a discontinuous jump at a finite frequency $\omega_c$. Due to the Kramers-Kronig relations, the real part exhibits a logarithmic singularity. Within RPA, Eq.~(\ref{eq:chi_s_sol}), this results in the spin resonance peak -- divergence of $\mathrm{Im}\chi_{+-}(\Q,\omega)$ at a frequency $\omega_R < \omega_c$. Frequency $\omega_c$ is determined by the two gaps, $\Delta_{\k \mu}$ and $\Delta_{\k+\Q \nu}$, `connected' by the wave vector $\Q$. If the gaps are equal, say $\Delta_L$, then the resonance peak appears at frequencies below $2\Delta_L$. If gaps are different and equal to $\Delta_L$ and $\Delta_R$, then the peak appears at $\omega_R \leq \Delta_L + \Delta_S$. Both these cases are shown in Figs.~\ref{fig:5orbImChi_set1} and~\ref{fig:5orbImChi_set2}.

To see the influence of the interaction parameters, we choose six sets of Hubbard $U$ and Hund's $J$ values. We also used the spin-rotational invariance constraint that minimizes the number of free parameters by setting $U'=U-2J$ and $J'=J$.
For the first three sets, the value of $U$ is chosen so that the system is near the magnetic instability; slight increase of $U$ results in the divergency of $\mathrm{Im}\chi$ at the wave vector $\Q$. Such choice is naturally related to the proximity of the system to the antiferromagnetic state at zero doping~\cite{Reviews}. The only parameter that we vary in this case is the Hund's exchange $J$. The other three sets are chosen to demonstrate what happens for smaller values of $U$ and the similar values of $J$.

In Fig.~\ref{fig:5orbImChi_set1}, results for $J=0$eV, $J=0.1$eV, and $J=0.15$eV with fixed $U=1.4$eV are shown. Note the increase of the spin response in all cases, both in normal and superconducting states. Spin resonance peak is shifted to lower frequencies. The energy scale $\omega_c$ stays the same because it is determined by the bare susceptibility, but the frequency at which $\mathrm{Im}\chi_{+-}(\Q,\omega)$ diverge changes and becomes smaller.

Similar situation is observed for smaller values of interaction parameters, see Fig.~\ref{fig:5orbImChi_set2} where results for $U=1.2$eV, $J=0$eV, $U=1.3$eV, $J=0$eV, and $U=1.3$eV, $J=0.1$eV are shown. For the smallest value of $U$, the resonance peak almost disappears, especially in the case of unequal gaps.
However, considering the fact that the slightly doped iron-based materials are antiferromagnets, the spin susceptibility should diverge at the nesting wave vector $\Q$ in the itinerant scenario for the magnetism. Therefore, one should expect the sizeable values of interaction parameters.

\section{Summary}

We studied the spin susceptibility in FeBS in the superconducting state with equal and unequal gaps, $\Delta_L$ and $\Delta_S$. Spin resonance appears in the $s_\pm$ state below the characteristic energy scale determined by the sum of gaps on two different Fermi surface sheets connected by the scattering wave vector $\Q$. We varied the interaction parameters, in particular, Hubbard repulsion $U$ and Hund's exchange $J$. With increase of interaction, we observe a total increase of the spin response both in normal and superconducting states. At the same time, the spin resonance peak is shifted to lower frequencies staying below the characteristic energy scale.

\begin{acknowledgements} This work was supported in part by the Russian Foundation for Basic Research (grant 16-02-00098), Presidium of RAS Program for the Fundamental Studies \#12, and ``BASIS'' Foundation for Development of Theoretical Physics and Mathematics. MMK acknowledges support by the Gosbudget program \# 0356-2017-0030.
\end{acknowledgements}

\end{document}